\documentclass[twocolumn,journal]{IEEEtran}

\usepackage{amsfonts}
\usepackage{amsmath}
\usepackage{amsthm}
\usepackage{amssymb}
\usepackage{graphicx}
\usepackage[T1]{fontenc}
\usepackage{supertabular}
\usepackage{longtable}
\usepackage[usenames,dvipsnames]{color}
\usepackage{bbm}
\usepackage{fancyhdr}
\usepackage{breqn}
\usepackage{fixltx2e}
\usepackage{capt-of}
\usepackage{tikz}
\usetikzlibrary{matrix}
\usepackage{endnotes}
\usepackage{soul}
\usepackage{marginnote}

\newcommand{\mathsym}[1]{}
\newcommand{\unicode}[1]{}

\usepackage{hyperref}

\usepackage{graphicx}
\graphicspath{{figures/}{/Users/nntaleb/Dropbox/Latex/SilentRisk/}}
 
\usepackage[framemethod=TikZ]{mdframed}
\usepackage[framemethod=TikZ]{mdframed}
\usepackage{framed}

\usepackage{enumitem}

\graphicspath{{figures/}{/Users/nntaleb/Dropbox/Latex/Collectednew/}}
\mdfsetup{%
nobreak=false,
   middlelinecolor=black,
   middlelinewidth=1pt,
   backgroundcolor=yellow!20,
   roundcorner=3pt}

\mdtheorem[nobreak=true,outerlinewidth=1,
backgroundcolor=yellow!50, outerlinecolor=black,innertopmargin=0pt,splittopskip=\topskip,skipbelow=\baselineskip, skipabove=\baselineskip,ntheorem,roundcorner=5pt,font=\itshape]{theorem}{Theorem}

\mdtheorem[nobreak=true,outerlinewidth=1,
backgroundcolor=yellow!20, outerlinecolor=black,innertopmargin=0pt,splittopskip=\topskip,skipbelow=\baselineskip, skipabove=\baselineskip,ntheorem,roundcorner=5pt,font=\itshape]{remark}{Comment}

\mdtheorem[nobreak=true,outerlinewidth=1,
backgroundcolor=pink!30, outerlinecolor=black,innertopmargin=0pt,splittopskip=\topskip,skipbelow=\baselineskip, skipabove=\baselineskip,ntheorem,roundcorner=5pt,font=\itshape]{principle}{Principle}

\mdtheorem[nobreak=true,outerlinewidth=1,
backgroundcolor=yellow!50, outerlinecolor=black,innertopmargin=5pt,splittopskip=\topskip,skipbelow=\baselineskip, skipabove=\baselineskip,ntheorem,roundcorner=5pt,font=\itshape]{background}{Background}

\usepackage{hyperref}
\title{{\color{Red}
Bitcoin, Currencies, and Fragility}}
\author{
    \IEEEauthorblockN{Nassim Nicholas Taleb\IEEEauthorrefmark{2}
    \IEEEauthorrefmark{3}     } \\
        \IEEEauthorblockA{\IEEEauthorrefmark{2}Universa Investments}\\
        \IEEEauthorblockA{\IEEEauthorrefmark{3}Tandon School of Engineering, New York University}\\  
      
       Forthcoming, \textit{Quantitative Finance}\\

   \thanks{NNT1@nyu.edu}
   \thanks{The author thanks Gur Huberman, Mark Spitznagel, Brandon Yarkin, Arthur Breitman, Trishank Karthik Kuppusamy, Jim Gatheral, Joe Norman, Zhuo Xi, David Boxenhorn,  Antonis Polemitis, Joe Shipman, and others for useful discussions.}}

\begin{document}
\maketitle

\begin{mdframed}

\section*{Introduction/Abstract}
This discussion applies quantitative finance methods and economic arguments to cryptocurrencies in general and bitcoin in particular ---as there are about $10,000$ cryptocurrencies, we focus (unless otherwise specified) on the most discussed crypto of those that claim to hew to the original protocol \cite{nakamoto2009bitcoin} and the one with, by far, the largest market capitalization. 

In its current version, in spite of the hype, bitcoin failed to satisfy the notion of "currency without government" (it proved to not even be a currency at all), can be neither a short nor long term store of value (its expected value is no higher than $0$), cannot operate as a reliable inflation hedge, and, worst of all, does not constitute, not even remotely, a safe haven for one's investments, a shield against government tyranny, or a tail protection vehicle for catastrophic episodes.

Furthermore, bitcoin promoters appear to conflate the success of a payment mechanism (as a decentralized mode of exchange), which so far has failed, with the speculative variations in the price of a zero-sum maximally fragile asset with massive negative externalities.

Going through monetary history, we show how a true numeraire must be one of minimum variance with respect to an arbitrary basket of goods and services, how gold and silver lost their inflation hedge status during the Hunt brothers squeeze in the late 1970s and what would be required from a true inflation hedged store of value.

\end{mdframed}

\begin{figure}
		\includegraphics[width=\linewidth]{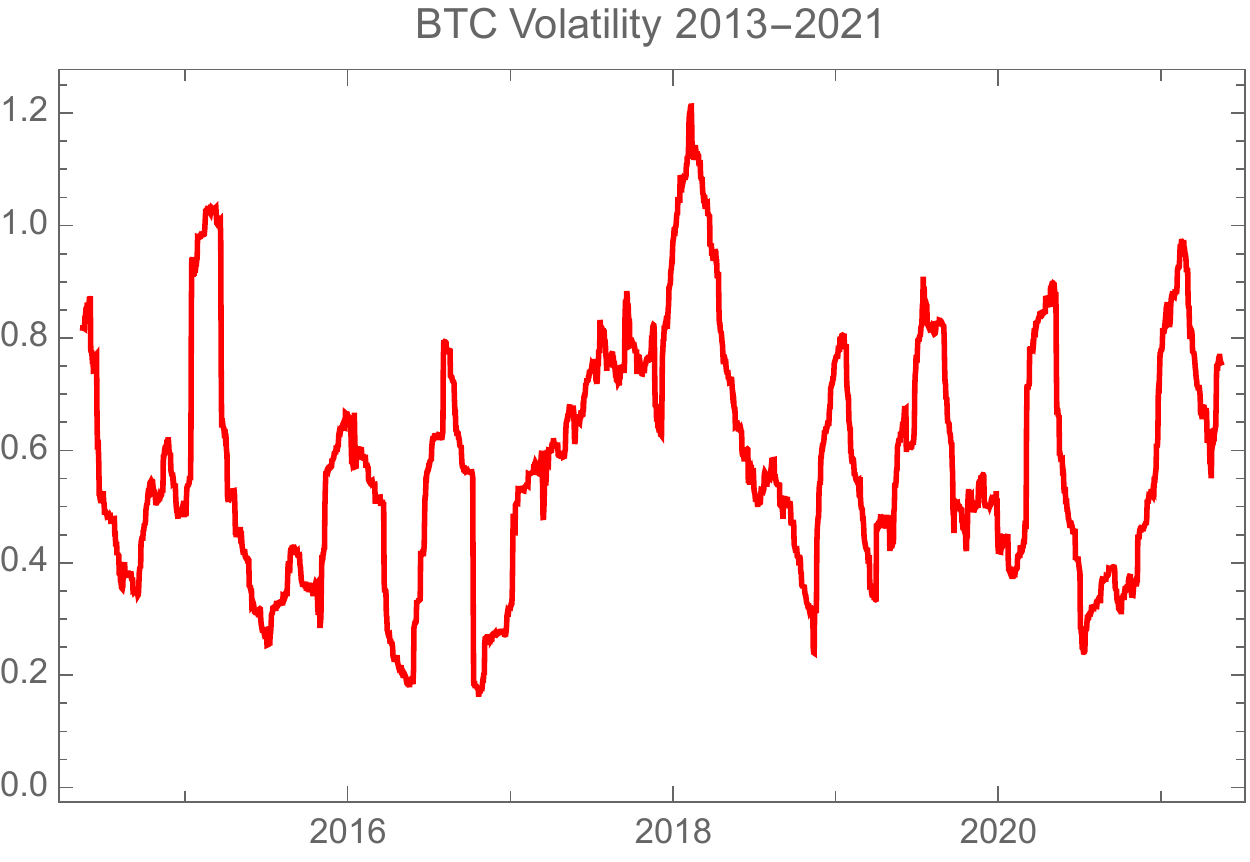}
\caption{BTC return, 3 months annualized volatility. It does not seem to drop over time. }\label{btcvolatility}
		\includegraphics[width=\linewidth]{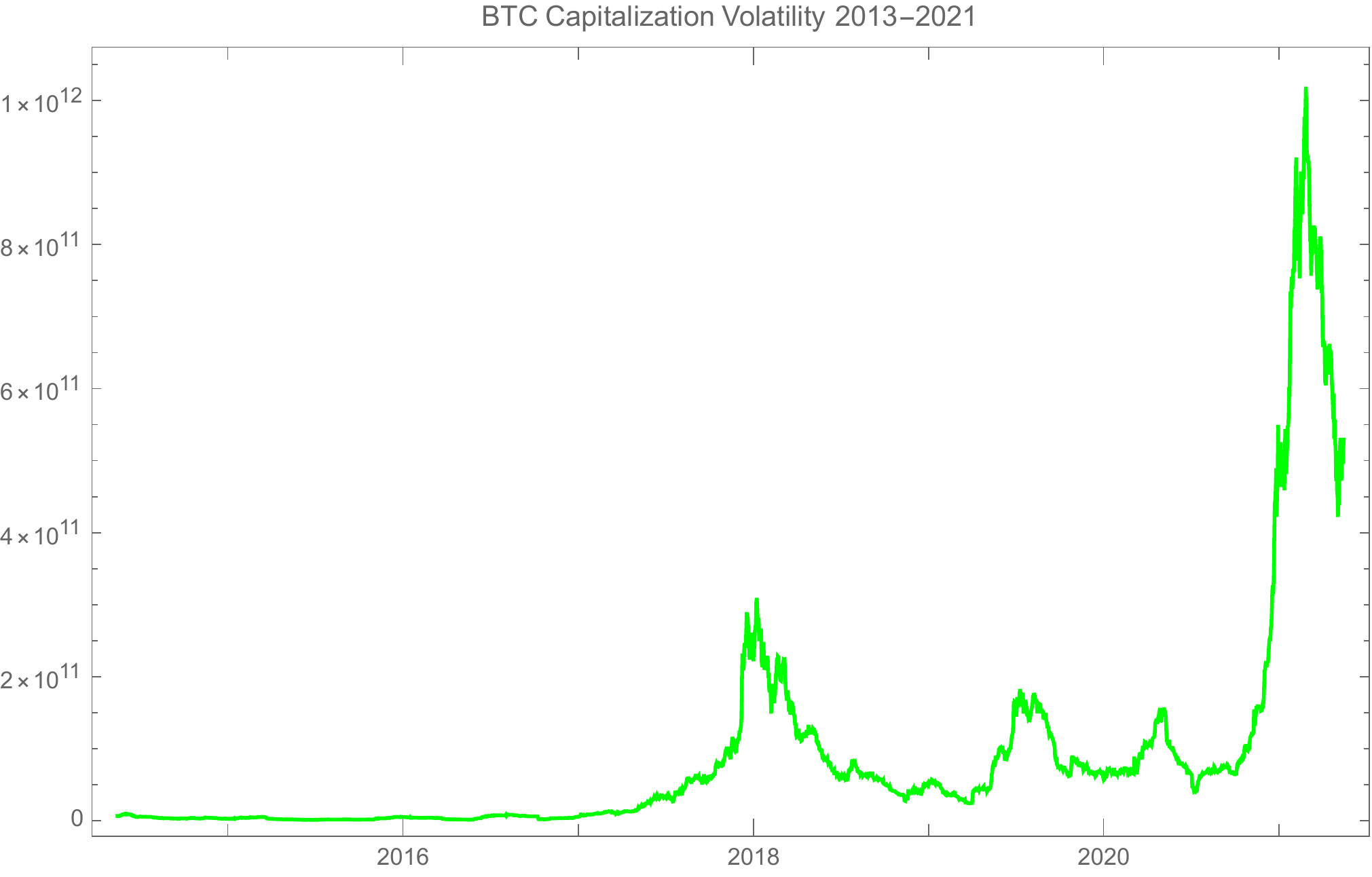}
\caption{Too volatile to fail? We show the volatility of the capitalization of BTC. At higher levels of capitalization, return volatility compounds. In 2021 a swing of half a trillion dollars in the capitalization of bitcoin took place.}\label{toovolatiletofail}
\end{figure}

\section*{The Blockchain}
First, let us consider what cryptocurrencies do by examining the notion of blockchain and its intellectual and mathematical appeal. 

The concept behind such a chain is quite intuitive to early practitioners of quantitative finance. Consider that before efficient software for Monte Carlo simulations became widely available, some of us were using methods to generate pseudorandom variables via some forms of chained nonlinear transformations, in the spirit of Von Neumann's original idea \cite{von195113}. Indexing sequences by $t= 1,2, \ldots n$, with a seed at $t$, a variable $x_t$ on the real line generates via nonlinear transformations $r: \mathbb{R} \to \mathbb{R}$, an output variable $r(x_t)$. This output variable can serve as a pseudorandom seed to generate another pseudorandom variable, $r(x_{t+1})$. For all $t$,  knowledge of  $r(X_t)$ allows knowledge of all subsequent variables $r(x_\tau)_{\tau > t}$  and replication of the entire sequence, thus probabilistically mimicking the arrow of time. It is also crucial that the same seed produces exactly the same pseudorandom variable, allowing  verification of sequence, but disallowing easy reverse engineering.

What the blockchain added, thanks to the hash function, is the condition that $r(.)$ must be functionally and probabilistically bijective: no two seeds should produce the same output (or should produce a vanishingly low probability of that happening), what, in computer science terminology, is called \textit{collision}.  

This hard-wired attribute and absence of supervision of the blockchain allow the storage of activities on a public ledger to facilitate peer-to-peer commerce, transactions, and settlements. The blockchain concept also allows for serial record keeping. This is supposed to help create what the original white paper \cite{nakamoto2009bitcoin} described as: 
\begin{quote}
A purely peer-to-peer version of electronic cash would allow online payments to be sent directly from one party to another without going through a financial institution.	
\end{quote}

From that paper, bitcoin makes use of three existing technologies: 1) the hash function, 2) the Merkle tree (to chain blocks of transactions tagged by the hash function), and 3) the concept of proof of work (used to deter spam by forcing agents to use computer time in order to qualify for a transaction) --- technologies that, ironically, all came out of the academic literature\cite{narayanan2017bitcoin}\footnote{As this discussion is focused on proof of work, we exclude from it Ethereum and other cryptocurrencies.}. The idea provides a game theoretic approach to mitigate the effects  of the absence of custodian and lack of trust between participants in the maintenance of a permanent shared public ledger --- attenuating or circumventing the coordination quandary known as the "Byzantine general problem". 

The bitcoin transactional currency (BTC) system establishes an adversarial collaboration between the so-called "miners" who validate transactions by getting them on a public ledger; as a reward they get coins  plus a fee from the underlying transactions, transfers of coins between parties. The proof of work method has an adjustable degree of difficulty based on the speed of blocks, which aims, in theory, to keep the incentive sufficiently high for miners to keep operating the system. Such adjustments lead to an exponential increase in computer power requirements, making at the time of writing onerous energy demands on the system --- energy that could find alternatives in  other computational and scientific uses.

Miners derive their compensation from both seignorage (the market value of a bitcoin minus its mining costs) and transaction fees upon validation --- with the plan to switch to transaction fees as the sole revenues upon the eventual depletion of the coins, which are limited to a fixed number. 

A central attribute is that bitcoin depends on the existence of such miners for perpetuity.

 Note that the entire ideological basis behind bitcoin is complete distrust of other operators --- there are no partial custodians; the system is fully distributed, though prone to concentration\footnote{From public data, we were able to verify that the distribution of holdings of bitcoin follows a powerlaw with tail index $\approx \frac{5}{4}$, no different from the distribution of wealth in the U.S.}.  Furthermore,  by the very nature of the blockchain, transactions are irreversible, no matter the reason.
  
Finally, note that bitcoins are zero-sum by virtue of the \textit{numerus clausus}.

As we will see, mathematical and combinatorial qualities do not necessarily translate into financial benefits at either individual or systemic levels.

\begin{remark}[Why BTC is worth exactly $0$]
Gold and other precious metals are largely maintenance free, do not degrade over an historical horizon, and do not require maintenance to refresh their physical properties over time.

 Cryptocurrencies require a sustained amount of interest in them.
\end{remark}

\section*{Vulnerability of revenue-free bubbles}

A central result (even principle) in the rational expectations and securities pricing literature is that, thanks to the law of iterated expectations, if we expect now \textit{that we will expect} the price to vary at some point in the future,  then by backward induction such a variation must be incorporated in the price \textit{now}. When there are no dividends, as with growth companies, there is still an expectation of future earnings, and a future expected reward to stockholders --- directly via dividends, or indirectly via reverse dilutions and buybacks.  It remains that a stock is a claim on accumulated assets and their residual value.

Earnings-free assets with no residual value are problematic.

The implication is that, owing to the absence of any explicit yield benefitting the holder of bitcoin, \textit{if} we expect that at any point in the future the value will be zero when miners are extinct, the technology becomes obsolete, or future generations get into other such "assets" and bitcoin loses its appeal for them, \textit{then} the value must be zero \textit{now}\footnote{Using a traditional rational bubble model (see \cite{blanchard1982bubbles} and the review by \cite{brunnermeier2016bubbles}), we get the following conditions. Let $r_d$ be a discount rate and $\pi$ be a probability of absorption over a period. To escape the barrier, bitcoin must grow at  $e^{r+\pi }$ \textit{forever}, but no more, without remission, and with total certainty. Should it grow then stabilize, it still would be prone to extinction. We note that traditionally, models rule out any continuous growth at an exponential rate faster than $r+\pi$ because the security or asset would then represent the entire economy. Bitcoin distinguishes itself from other assets because of its fragility as a mere book entry on a virtual ledger that requires constant refreshing \textit{ad infinitum}.}.  

The typical comparison of bitcoin to gold is lacking in elementary financial rigor\footnote{It is also a reasoning error to claim that an innovation, bitcoin, can become the "new gold" \textit{ab ovo}, when gold wasn't decided to be so by fiat thanks to a white paper; it organically became a reserve asset ex post, through centuries of competitive selection against other modes of storage, payment, and collectibles. Gold elicited an aesthetic fascination and had been used as jewelry and store of value for more than two millennia before it became, literally, a currency or before there was such a thing as a currency. The Phoenicians used it as store of value because there was demand for it, and it was not until the $6^{th}$ C. BCE that coins from Sardis became a widespread means of exchange \cite{graeber2012debt}.}.
{We will see below how precious metals lost their quality as a medium of exchange; gold and other dividend-free precious items (such as other metals or stones) have held some financial status for more than $6,000$ years, and their physical status for several orders of magnitude longer (i.e., they did not degrade or mutate into some other alloy or mineral). So one can expect one's gold or silver possessions to be around physically for at least the next millennium, as well as  having some residual economic value by iteration, for the same reason. Metals have ample industrial uses with demand elasticity (and substitution for other raw materials). Currently, about half of  gold production goes to jewelry (for which there are often no storage costs), one tenth to industry, and a quarter to central bank reserves.

Path dependence is a problem. We cannot expect a book entry on a ledger that requires active maintenance by interested and incentivized people to keep its \textit{physical} presence, a condition for monetary value,  for any  period of time --- and of course we are not sure of the interests, mindsets, and preferences of future generations. Once bitcoin drops below a certain threshold, it may hit an absorbing barrier and stays at 0 --- gold  on the other hand is \textit{not} path dependent in its physical properties\footnote{The absorbing barrier does not have to be $0$ for the price to spiral to $0$ upon hitting the barrier. This is similar to saying "if the heart rate drops below ten beats per minutes, it will be $0$ (death)" --- nor does it necessarily have to be caused by a drop in price. Nor does it have to be endogenous.}. As discussed in \cite{taleb2012antifragile}, technologies tend to be supplanted by other technologies with a vulnerability in proportion to their past survival duration (>99\% of the new is replaced by something newer), whereas items such as gold and silver have proved resistant to extinction}. Furthermore bitcoin is supposed to be hacker-proof and is based on total infallibility in the future, not just at present. It is crucial that bitcoin is based on perfect immortality; unlike  conventional assets, the slightest mortality rate puts its value at $0$\footnote{To counter the effect of the absorbing barrier, the asset must grow at an exponential rate \textit{forever}, without remission, and with total certainty. \\
Belief in such an immortality for BTC --- and its total infallibility --- is in line with the common observation that its enthusiastic investors have the attributes of a religious cult.}.

\begin{principle}[Cumulative ruin]
If any non-dividend yielding asset has the tiniest  probability of hitting an absorbing barrier (causing its value to become $0$), then its present value must be $0$.
	\end{principle}

We exclude collectibles from that category, as they have an aesthetic utility as if one were, in a way, renting them for an expense that maps to a dividend --- and thus are no different from perishable consumer goods. The same applies to the jewelry side of gold: my gold necklace may be worth $0$ in thirty years, but then I would have been wearing it for six decades.

The difference between the current bitcoin bubble and past recent ones, such as the dot-com episode spanning the period over 1995-2000, is that shell companies were at least promising some type of future revenue stream. Bitcoin would be allowed to escape a valuation methodology had it proven to be a medium of exchange or satisfied the condition for a numeraire from which other goods could be priced. But currently it is not, as we will see next.

\section*{Success in wrong places}

More generally, the fundamental flaw and contradiction at the base of most cryptocurrencies is, as we saw, that the originators, miners, and maintainers of the system currently make their money from the inflation of their currencies rather than \textit{just} from the volume of underlying transactions in them. Hence the total failure of bitcoin to become a currency has been masked by the inflation of the currency value, generating (paper) profits for a large enough number of people to enter the discourse well ahead of its utility. 

\begin{remark} [Success for a digital currency]
There is a mistaken conflation between success for a "digital currency", which requires some stability and usability, and speculative price appreciation.\label{success}
\end{remark}

Transactions in bitcoin are considerably more expensive than wire services or other modes of transfers, or ones in other cryptocurrencies\footnote{Transactions in bitcoin are orders of magnitude more expensive than those done using African mobile phones.}. They are order of magnitudes slower than standard commercial systems used by credit card companies ---anecdotally, while you can instantly buy a cup of coffee with your cell phone, you would need to wait ten minutes if you used bitcoin\footnote{"As it grew in popularity, Bitcoin became cumbersome, slow, and expensive to use. It takes about 10 minutes to validate most transactions using the cryptocurrency and the transaction fee has been at a median of about \$20 this year." By Eswar Prasad, \textit{New York Times}, Jun 15, 2021.}. They cannot compete with African mobile money.
 \footnote{There appear to be other protocols issued from the original white paper that claim to be more transaction focused; as with Ethereum, we exclude them from this analysis.}. Nor can the system outlined above ---as per its very structure ---accommodate a large volume of transactions --- which is something central for such an ambitious payment system.

To date, twelve years into its life, in spite of all the fanfare, but with the possible exception of the price tag of Salvadoran permanent residence (3 bitcoins), there are currently no prices fixed in bitcoin floating in fiat currencies in the economy.

\begin{figure}
		\includegraphics[width=\linewidth]{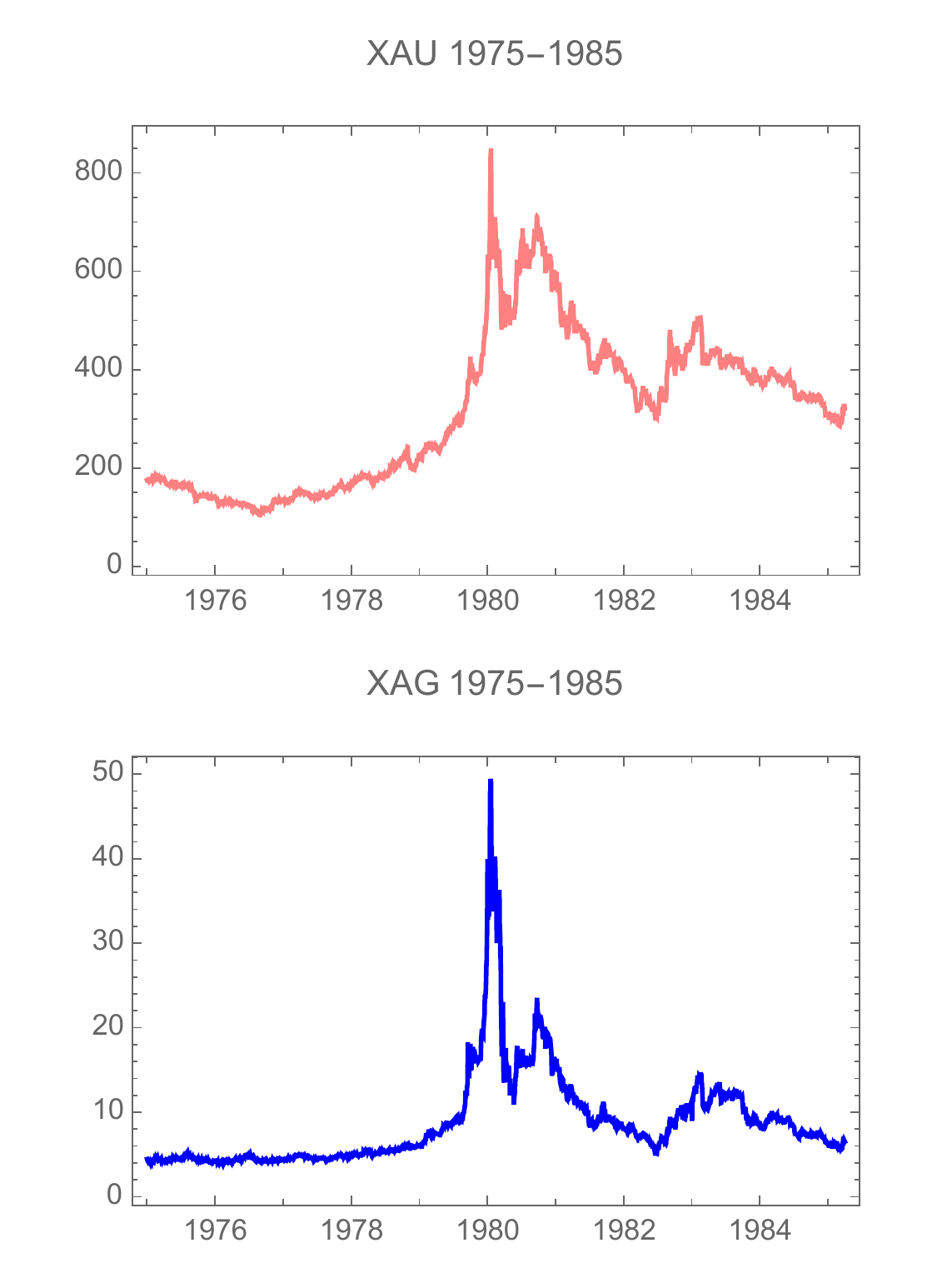}
\caption{The rise and fall of metals during the Hunt squeeze of silver and, indirectly, gold.}\label{hunt}
\end{figure}

\section*{Principles for a currency}
First, let's discuss the demonetization of gold. In 1971, the U.S. government terminated the Bretton Woods Agreement, ending the convertibility of the U.S. dollar into gold. Gold stocks were growing 
too slowly, and, as  mentioned earlier, much of it went to jewelry and industry --- the most robust theory is that there was not enough gold to keep up with economic growth\footnote{ Ironically the U.S. deficit caused the dollar to be more widely available and used, in stable supply, by what is called the Triffin paradox.}. Furthermore, there had been long debates over the hampering of monetary policy by sticking to metals, as witnessed by the bullionist controversy\footnote{Even Ricardo got drawn in, see Ricardo's 1811-1816 arguments \cite{ricardo1811reply},\cite{ricardo1816proposals}, and commentary by Jevons \cite{jevons1863serious}. }. It appears that developed economies have trouble hooking their currencies to a commodity.

In the early 1970s, the Hunt brothers started to hoard silver (when they started, U.S. citizens were banned from directly owning gold),  and accelerated their hoarding in the late 1970s, turning it into a squeeze. It lead to a speculative explosion in the price of silver, as shown in Fig \ref{hunt}, leading by contagion to between a fivefold and tenfold increase in the price of precious metals.  Then, upon the deflation of the bubble, metals gave back more than half of their gains and languished for more than two decades. At the time of writing, 41 years later, neither gold nor silver have, inflation adjusted, reached their previous peak. The same effect took place in 2008-2009 in the wake of the banking crisis: gold and silver jumped upwards between 80 and 120 \% then subsequently lost most of their gains. 

 Gold and silver proved then that they could neither be a reliable numeraire, nor an inflation hedge. The world had become too sophisticated for precious metals. If we consider the most effective numeraire, it must be the one in which the bulk of salaries are paid, as we will show next. 

\begin{remark}[Payment system]
There is a conflation between "accepting bitcoin for payments" and pricing goods in bitcoin. To "price" in bitcoin, bitcoin the price must be fixed, with a conversion into fiat floating, rather than the reverse.
\end{remark}


Let us go deeper into how a currency can come about. No transaction between two persons is analytically pairwise in an open economy.	The root of the confusion lies in the prevalent na\"ive-libertarian illusion that a transaction between two consenting adults, when devoid of coercion, is effectively just a transaction between two consenting adults and can be isolated and discussed as such\footnote{www.libertarianism.org}. 
But one must consider the ensemble of transactions and the interactions between agents: people happen to engage in contractual agreements with others; for them a specific transaction is just one piece. To be able to regularly buy  goods denominated in bitcoin (whose prices fixed in bitcoin but floating in U.S.\$ or some other fiat currency), one must have an income that is fixed in bitcoin. Such an income must come from somewhere, say, an employer. For an employer to pay a salary fixed in bitcoin, she or he must be getting revenues fixed in bitcoin.  Furthermore, for the vendor to offer a can of beer in fixed bitcoins, she or he must be paying for the raw material, and have the overhead fixed in bitcoin. The same applies to the mismatch of assets and obligations on a balance sheet. All this requires a parity in bitcoin-USD of low enough volatility to be tolerable and for variations to remain inconsequential.

There are also arbitrage bounds present in any sufficiently efficient economy with relatively free markets. 

Furthermore, if a vendor prices goods in bitcoin, and the value fluctuates from the initial fixing, the price will be directly or indirectly arbitraged: when the conversion rate to fiat is favorable, customers will buy from the bitcoiner; when it is unfavorable they will either buy elsewhere (indirect arbitrage), or if possible, return previously purchased goods (direct arbitrage). For the price to not be arbitrageable requires the good to be unique and unavailable elsewhere at a price fixed in another currency --in this case it becomes, simply, a proxy for bitcoin. The only items that currently appear to be somewhat priced in bitcoin are other cryptocurrencies, even then not always.

Bimetalism did not last long \cite{velde2000model}, nor could commodities last as currencies in developed economies\cite{sargent1983model}. 

More generally, the reasons multiple currencies exist (in the absence of pegs) is because there is not enough globalization and markets are not entirely free between currency zones. And some goods and services,  "such as haircuts and auto repair cannot be traded internationally" \cite{krugman2017international} ; they are not, to use the language of quantitative finance, arbitrageable.

In 2021, the governments (central and local) share of GDP in Western economies is around 30-60\%, one order of magnitude higher than it was in the 1900s. Government employees and contractors get paid in fiat currency; taxes are collected similarly \footnote{The designation "fiat" is a misleading stretch of language: money is not created by edict but largely via credit, by governments or the private sectors, particularly the banking system --- and both lenders and borrowers need the least volatile currency \cite{mcleay2014money}.}. 

Finally, while within a modern currency zone a bimetallic style dual currency cannot easily exist, the same limitations exist between currency zones; parity between currencies  tend to be subjected to volatility bounds. An observation we currency option traders made while doing cross-currency volatility arbitrages is that the volatility of a currency pair is inversely proportional to the trade between the two currency zones --- countries heavy into trade such as Hong Kong, Saudi Arabia, the UAE, and Singapore (at some point) have maintained explicit pegs to the U.S. dollar or some basket. There could be an interactive relationship between trade and volatility: one can argue that the stability of a currency-pair (adjusted for the yield curve) encourages trade and trade in turn brings stability to the pair\footnote{Currency pairs often show fake volatility as the spot price can be fluctuating, but forward contracts do less so, owing to interest rate adjustments in the weak currency: interest rates rise to compensate holders for the devaluation.}$^{,}$\footnote{ We note here that quantitative finance operates along the lines of neoclassical economic theory in that both share a central principle: absence of arbitrage, which maps to the law of one price --- the former, a concept initially aimed at goods and services, may be broadened to include asset valuation \cite{ross2009neoclassical}. When we apply the law of one price to currencies, we realize using basic arbitrage arguments that the recent globalization does not allow for different currencies to coexist in the same marke: one must win.}.

Now bitcoin, as seen in Fig.\ref{btcvolatility} has maintained extremely high volatility throughout its life (between 60\% and 100\% annualized) and, what is worse, at higher prices, which makes it's capitalization considerably more volatile, rising in price as shown in Fig. \ref{toovolatiletofail} --- is it too volatile to fail?
\section*{The difficulty with inflation hedges}

This does not mean that a cryptocurrency cannot displace fiat --it is indeed desirable to have at least one \textit{real} currency without a government. But the new currency just needs to be more appealing as a store of value by tracking a weighted basket of goods and services with minimum error.

Displacing fiat is not easy, and has been done locally ---though no single item has proved to be permanent and the difficulty is best represented in the following example. During the 1970s, the Italian national telephone tokens, the \textit{gettoni}, were considered acceptable tender, almost always accepted as payment.  
The price of the espresso when expressed in lira varied over time, but it remained sticky to the \textit{gettone}. For a while the \textit{gettone} proved the closest money to track the Fisher Index across 12 communes\cite{campiglio1986analisi}\footnote{Likewise, the M-Pesa mobile currency used as tender in Africa is associated with transferable airtime minutes \cite{nair2018electronic}. People can do microfinance via cell phones.}. And while the gettoni worked for daily purchases such as espresso, it is doubtful that they could have been used as payment for an Alfa Romeo \cite {coluccifenomeno}.

Considering that communications get cheaper over time, the notion of a telephone call is today, in the Zoom days, obsolete. So the \textit{gettone} story illustrates the fact that, owing to technological changes, in the long term, no single item, such a telephone call, will  permanently track inflation indices and act as a store of value. Even categories have their weights naturally revised over time: the share of food and clothing  declined by almost threefold as a proportion of  Western consumers expenditure since the great recession.

Thus we can look at an inflation hedge as the analog of a minimum variance numeraire.
 \begin{mdframed}
	\textit{Let us assume that there exists an efficient inflation hedge for period $[t_0, T]$  for an index methodology, the one in which the index, constantly revised, is the most stable when it is as a numeraire (adjusting for interest and dividend payments).}
 \end{mdframed}
Can one find her or his own hedge? 

In the parable of the Christ in the temple, Jesus kicked the money changers out of the temple of Jerusalem... Now one wonders why were there were money changers in a place of worship? The answer is that the temple took for currency only the shekel of Tyre, known for its 90\% silver content and its ancestral quality control \cite{murphy2000jesus}\footnote{This appears to be a Judean custom; in
the Mishnah (Bekhorot 8): "The five sela coins of the redemption of the firstborn son, with regard to which it is written: "Five shekels of silver, after the shekel of the Sanctuary" (Numbers 18:16), are calculated using a Tyrian maneh. The silver content of the Tyrian coinage is significantly higher than that of provincial coinage, which is worth one-eighth its value."}. 

Simply, there is a free market for fiat currencies, with the most reliable \textit{at the time} used by third parties.  
Before the Euro, there were plenty of currencies in Europe. But long term contracts, investments, and commitments were evaluated in deutschmarks or Swiss francs, sometimes the U.S. dollar; drachmas, liras, and pesetas were there mostly for petty expenditures. So what we had was competition between fiat currencies just as with the shekel-of-Tyre! 

This competition provides for a vastly more convenient monetary store of value. For practitioners of quant finance, the most effective inflation hedge can be a combination of bets which includes the short bond.

\section*{Some additional fallacies}

\subsubsection{Fallacy of libertarianism}  The belief that bitcoin is an offshoot of libertarian and Austrian economics has no solid backing --- it has the same lack of rigor as the one behind the belief that cryptos represent a "hedge for inflation".  Spitznagel \cite{spitz2017} had already, in 2017, debunked the notion that bitcoin can be a safe haven (as discussed next) or that the principles of Austrian economics can be invoked in support of cryptocurrencies.

\begin{remark}[Law vs. Regulations vs. Rules]
	Libertarianism is about the rule of law in place of the rule of regulation. It is not about the rule of rules.
\end{remark}

Libertarianism is fundamentally about the rule of law in place of the rule of regulation. It is not about the rule of rules ---  mechanistic, automated rules with irreversible outcomes.  The real world is fraught with ambiguities and even Napoleonic law (far less mechanistic than crypto rules) cannot keep up --- to wit, as a risk management directive, most commercial contracts traditionally prefer forums of dispute resolution to be under the more flexible Anglo Saxon common law (London, NY, Hong Kong) that rules on balance, intent, and symmetry in contracts. This applies of course to quantitative finance products such as complex derivatives contracts for which one needs to minimize the legal risk.

Nor is libertarianism about total distrust. 

\subsubsection{Fallacy of safe haven, I (protection for financial tail risk)} The experience of March 2020, during the market panic upon the onset of the pandemic, when bitcoin dropped farther than the stock market ---and subsequently recovered with it upon the massive injection of liquidity is sufficient evidence that it cannot remotely be used as a tail hedge against systemic risk. Furthermore, bitcoin appears to respond to liquidity, exactly like other bubble items.

It is also uncertain what could happen should the internet experience a general, or an even a regional, outage --- particularly if it takes place during a financial collapse.

\subsubsection{Fallacy of safe haven, II (protection from tyrannical regimes)}
To many paranoid antigovernment individuals and of others distrustful of institutions, bitcoin has been marketed as a safe haven --- also with an open invitation to fall for the fallacy that a volatile electronic token in a public setting is a place for your hidden treasure. 

By its very nature, bitcoin is open for all to see. The belief in one's ability to hide one's assets from the government with a public blockchain easily triangularizable at endpoints, and not just read by the FBI but also by people in their living rooms, requires a certain lack of financial seasoning and statistical understanding --- perhaps even a lack of minimal common sense. For instance a Wolfram Research specialist was able to statistically detect and triangularize "anonymous" ransom payments made by Colonial Pipeline on May 8  in 2021 \cite{porechna2021} --- and it did not take long for the FBI to restore the funds.

We can safely assume that government structures and computational power will remain stronger than those of distributed operators who, while distrusting one another, can fall prey to simple hoaxes.

 In the cyber world, connections are with people one has never met in real life; infiltration by government agents has proven to be extremely easy\footnote{This is one of the weaknesses of total decentralization.}. By comparison, the mafia required a Sicilian lineage for "friends of ours" for security clearance.  One never knows the degree of governmental surveillance and its real capabilities. 

The slogan "Escape government tyranny \textit{hence} bitcoin" is similar to advertisements in the 1960s extolling the health benefits of cigarettes.

\subsubsection{  Fallacy of the Agency problem}
One might have the impression that, by being distributed, Bitcoin would be democratic and reduce the agency problem perceived to be present among civil servants and bankers. Unfortunately, there appears to be a worse agency problem: a concentration of insiders hoarding what they think will be the world currency, so others would have to go to them later on for supply. They would be cumulatively earning trillions, with many billionaire "Hodlers" --- in comparison the "evil civil servants" behind fiat money make, at best, lower middle class wages. This situation represents a wealth transfer to the cartel of early bitcoin accumulators.

\section*{Conclusion}

We have presented the attributes of the blockchain in general and bitcoin in particular. Few assets in financial history have been more fragile than bitcoin.

The customary standard argument is that "bitcoin has its flaws but we are getting a great technology; we will do wonders with the blockchain". No, there is no evidence that we are getting a great technology --- unless "great technology" doesn't mean "useful". And  at the time of writing ---in spite of all the fanfare --- we have done still close to nothing with the blockchain.

 So we close with a Damascus joke. One vendor was selling the exact same variety of cucumbers at two different prices. "Why is this one twice the price?", the merchant was asked. "They came on higher quality mules" was the answer. 
 
 We only judge a technology by how it solves problems, not by what technological attributes it has.

\bibliographystyle{IEEEtran}
\bibliography{/Users/nntaleb/Dropbox/Central-bibliography}
%
%
%
%
%
%
%


\end{document}